\documentclass{article}

\usepackage{arxiv}
\usepackage[utf8]{inputenc} 
\usepackage[T1]{fontenc}    
\usepackage{nicefrac}       
\usepackage{microtype}      
\usepackage{amsmath}
\usepackage{amssymb}
\usepackage{amsfonts}
\usepackage{amsthm}         
\usepackage[shortcuts,acronym]{glossaries}

\usepackage{graphicx}
\usepackage{booktabs}       
\usepackage{multirow}
\usepackage[table]{xcolor}
\usepackage[caption=false,font=normalsize,labelfont=sf,textfont=sf]{subfig}

\usepackage[numbers]{natbib}
\usepackage{doi}
\usepackage{multirow}       
\usepackage{algorithm}      
\usepackage{algorithmic}    
\usepackage{soul}

\newacronym{IoT}{IoT}{Internet of Things}
\newacronym{MnL}{MnL}{Manifold Learning}
\newacronym{ML}{ML}{Machine Learning}
\newacronym{DL}{DL}{Deep Learning}
\newacronym{AI}{AI}{Artificial Intelligence}
\newacronym{FCNN}{FCNN}{Fully Connected Neural Network}
\newacronym{AE}{AE}{Autoencoder}
\newacronym{P-UMAP}{P-UMAP}{Parametric Uniform Manifold Approximation and Projection}
\newacronym{UMAP}{UMAP}{Uniform Manifold Approximation and Projection}
\newacronym{MSE}{MSE}{Mean Square Error}
\newacronym{GNN}{GNN}{Graph Neural Network}
\newacronym{GNNs}{GNNs}{Graph Neural Networks}
\newacronym{GIN}{GIN}{Graph Isomorphism Network}
\newacronym{NIDS}{NIDS}{Network Intrusion Detection Systems}
\newacronym{XAI}{XAI}{Explainable Artificial Intelligence}
\newacronym{SHAP}{SHAP}{Shapley Additive Explanations}
\newacronym{LIME}{LIME}{Local Interpretable Model-agnostic Explanations}
\newacronym{MLP}{MLP}{Multi-Layer Perceptron}
\newacronym{ReLU}{ReLU}{Rectified Linear Unit}
\newacronym{DDoS}{DDoS}{Distributed Denial of Service}
\newacronym{DoS}{DoS}{Denial of Service}
\newacronym{GNN-AE}{GNN-AE}{Graph Neural Network Autoencoder}
\newacronym{GNN-CLS}{GNN-CLS}{Graph Neural Network Classifier}
\newacronym{DBI}{DBI}{Davies-Bouldin Index}

\title{Interpreting Manifolds and Graph Neural Embeddings from Internet of Things Traffic Flows}

\date{\today}

\newif\ifuniqueAffiliation
\uniqueAffiliationfalse

\usepackage{authblk}

\newbox{\orcid}\sbox{\orcid}{\includegraphics[scale=0.06]{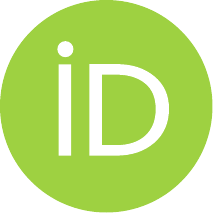}} 

\author[1]{%
    \href{https://orcid.org/0009-0005-5068-3166}{\usebox{\orcid}\hspace{1mm}Enrique Feito-Casares\thanks{\texttt{enrique.feito@urjc.es}}}%
}
\author[1]{%
    \href{https://orcid.org/0000-0001-6916-6082}{\usebox{\orcid}\hspace{1mm}Francisco M. Melgarejo-Meseguer}%
}

\author[2]{%
	\href{https://orcid.org/0000-0003-2024-7572}{\usebox{\orcid}\hspace{1mm}Elena Casiraghi}%
}
\author[2]{%
	\href{https://orcid.org/0000-0002-5694-3919}{\usebox{\orcid}\hspace{1mm}Giorgio Valentini}%
}

\author[1,3]{%
    \href{https://orcid.org/0000-0003-0426-8912}{\usebox{\orcid}\hspace{1mm}José-Luis Rojo-Álvarez}%
}

\affil[1]{Universidad Rey Juan Carlos, Madrid, Spain}
\affil[2]{Università degli Studi di Milano, Milan, Italy}
\affil[3]{Centro de Investigación for Data, Complex Networks and Cybersecurity Sciences,  \protect\\ Universidad Rey Juan Carlos, Madrid, Spain.}

\renewcommand{\shorttitle}{Interpreting Manifolds and Graph Neural Embeddings from Internet of Things Traffic Flows}

\hypersetup{
pdftitle={Interpreting Manifolds and Graph Neural Embeddings from Internet of Things Traffic Flows},
pdfsubject={cs.CR, cs.LG, cs.NI},
pdfkeywords={Graph Neural Networks, IoT Networks, Network Topology, Interpretable AI, Manifold Learning},
}

\newcommand\blfootnote[1]{
    \begingroup
    \renewcommand\thefootnote{}\footnote{#1}
    \addtocounter{footnote}{-1}
    \endgroup
}

\begin{document}
\maketitle

\begin{abstract}
The rapid expansion of Internet of Things (IoT) ecosystems has led to increasingly complex and heterogeneous network topologies. Traditional network monitoring and visualization tools rely on aggregated metrics or static representations, which fail to capture the evolving relationships and structural dependencies between devices. Although Graph Neural Networks (GNNs) offer a powerful way to learn from relational data, their internal representations often remain opaque and difficult to interpret for security-critical operations. Consequently, this work introduces an interpretable pipeline that generates directly visualizable low-dimensional representations by mapping high-dimensional embeddings onto a latent manifold. This projection enables the interpretable monitoring and interoperability of evolving network states, while integrated feature attribution techniques decode the specific characteristics shaping the manifold structure. The framework achieves a classification F1-score of 0.830 for intrusion detection while also highlighting phenomena such as concept drift. Ultimately, the presented approach bridges the gap between high-dimensional GNN embeddings and human-understandable network behavior, offering new insights for network administrators and security analysts.
\end{abstract}


\blfootnote{This work has been submitted to the IEEE for possible publication}
\section{Introduction}
\label{sec:introduction}

The rapid expansion of  \ac{IoT} ecosystems has fundamentally transformed the digital landscape, leading to increasingly complex and heterogeneous network topologies. Despite the benefits of cost-effective deployment and ubiquity, this diversity creates significant security vulnerabilities that expose the network to potential cyber threats. In this context, \ac{NIDS} serve as critical defense mechanisms, yet the evolving nature of \ac{IoT} traffic renders conventional monitoring approaches insufficient \cite{Xu2023}. Specifically, traditional network monitoring tools typically rely on tabular data, linear log streams, or aggregated metric plots, which inherently limit visualization and in-depth analysis capabilities.

Since network traffic is fundamentally a graph of interconnected devices rather than a simple list of events, \acp{GNN} have emerged as a natural architecture for this domain \cite{Sun2025}. By directly encoding the network topology, \acp{GNN} capture structural dependencies that tabular-based methods fail to capture. However, as with other \ac{DL} paradigms, this architecture presents a significant trade-off regarding interpretability. The high-dimensional embeddings generated by these models operate as a black box, establishing an understanding gap between the hidden latent spaces and human-interpretable network behaviors and this opacity limits the compliance and trust required for security-critical operations \cite{Zhou2022a}.

To bridge this interpretability gap, we adopt \ac{MnL} as a principled mechanism to translate high-dimensional \ac{GNN} representations into directly visualizable low-dimensional structures \cite{Fefferman2016,Ghojogh2023a}. \ac{MnL} is grounded in the assumption that high-dimensional observations lie on a low-dimensional latent manifold capturing the data intrinsic degrees of freedom. When learned from network-traffic representations, the resulting manifold provides a compact characterization of network states: benign traffic is expected to form stable neighborhoods, whereas deviations from the learned structure can indicate anomalous or malicious behavior. In this way, the manifold acts as an intermediate representation linking abstract embeddings to security-relevant network dynamics, enabling monitoring and interoperability under evolving \ac{IoT} conditions.

Building on this rationale, we introduce an \ac{MnL}-based framework for \ac{IoT}  network-traffic analysis that integrates representation learning and interpretability within a single training process. Rather than treating prediction and interpretation as decoupled stages, we train a \ac{GNN} using a joint objective that combines an intrusion-detection loss with a \ac{UMAP}-based manifold-preserving loss, thereby enforcing alignment between the internal embeddings and their low-dimensional manifold representation. By making the low dimensional projection an explicit optimization target, this coupling mitigates a key limitation of post-hoc visualization, where projections can drift away from the decision-relevant features used by the model. Finally, we complement the visual interpretability of the latent space with feature attribution via \ac{SHAP} \cite{Shapley1953,Roth1988}, therefore providing a dual-layered explanation handling both visual and numerical interpretability representations derived from the same learning process.

The remainder of this paper is organized as follows: Section~\ref{sec:related-work} reviews work related to the contributions of the paper. Section~\ref{sec:algorithms} describes the foundational algorithms and architectures employed in our approach. Section~\ref{sec:dataset} provides a detailed description of the  \ac{IoT} network dataset used for evaluation. Section~\ref{sec:methodology} presents our proposed methodology, including graph construction,  \ac{GNN} architecture, manifold projection, and explainability techniques. Section~\ref{sec:experiments} presents comprehensive experimental results and analysis. Section~\ref{sec:discussion} discusses implications, limitations, and future research directions.
\section{Related Work} \label{sec:related-work}

The paradigm of network traffic analysis has shifted significantly from tabular, flow-based heuristics to structural representation learning. As detailed in foundational literature, the field has moved toward modeling network entities as graph structures to preserve relational semantics. Early works focused on the transformation of raw logs into navigable topologies for forensic purposes. For instance, the Sec2graph approach by Leichtnam et al.~\cite{Leichtnam2020} and the GRANEF toolkit by Čermák and Šrámková~\cite{Cermak2021} proposed modeling security objects such as IP addresses and ports as nodes, with their interactions forming the edges. To address the temporal aspect of network flows, Apruzzese et al.~\cite{Apruzzese2020} introduced temporal graph representations for NetFlow data, explicitly capturing the time causality required to trace lateral movement and pivoting attacks. However, these earlier systems were primarily designed for data storage and manual forensic traversal rather than for automated feature learning.

Recent comprehensive surveys have highlighted the primary research trajectories and remaining challenges in this domain. Lagraa et al.~\cite{Lagraa2024} identifed a fundamental dichotomy in graph construction strategies, distinguishing between flow-centric graphs which offer granular detail but high computational overhead and host-centric aggregations that improve scalability at the cost of losing temporal resolution. In their extensive survey of \acp{GNN} for intrusion detection, Bilot et al.~\cite{Bilot2023} highlighted that while Message Passing Neural Networks have achieved state-of-the-art detection performance, a significant barrier to their operational deployment is the black-box nature of the learned representations. Regarding the evolution of the field, Dong et al.~\cite{Dong2023} emphasized that next-generation \ac{IoT} imposes unprecedented heterogeneity constraints, arguing that as systems evolve from homogeneous to complex multi-modal environments, traditional models struggle with data fusion and resource efficiency on edge devices. Complementing this view, Tung et al.~\cite{Tung2025} contended that standard \acp{GNN} face difficulties generalizing across the massive scale of these networks and explicitly propose the use of EXIT charts (Extrinsic Information Transfer charts) to diagnose information flow, thereby rendering the latent reasoning accessible to human analysts. Collectively, these reviews point toward a critical research gap, the need for architectures that can manage dynamic, heterogeneous topologies while ensuring both scalability and, above all, interpretability.

With the integration of \ac{DL} to address these gaps, research efforts pivoted toward utilizing  \acp{GNN} for automated anomaly detection. Gioacchini et al.~\cite{Gioacchini2023} demonstrated the efficacy of dynamic embeddings by employing Temporal Graph Attention Networks on Darknet traffic, generating 128-dimensional representations that successfully characterize evolving threats where static baselines fail. Similarly, Abou Rida et al.~\cite{AbouRida2022} investigated inductive learning capabilities using Convolutional  \acp{GNN} (specifically GraphSAGE), achieving robust detection rates with compact 64-dimensional embeddings. While these approaches validate the ability of  \acp{GNN} to reduce reliance on manual feature engineering, they predominantly focus on  closed-set classification metrics, often overlooking the geometric continuity of the latent space which is essential for identifying out-of-distribution anomalies.

In the specific domain of \ac{IoT}, where device heterogeneity necessitates specialized strategies, recent works have diversified into specific detection tasks. Gao et al.~\cite{Gao2023} addressed the core challenge of anomaly traffic detection by developing a  \ac{GNN} framework that aggregates neighbor features to identified deviations in \ac{IoT} communication patterns. Addressing the complexity of multi-stage intrusions, Altaf et al.~\cite{Altaf2024} proposed a  \ac{GNN}-based analysis specifically designed for the detection of sequential attacks, utilizing graph structures to correlate events across time steps rather than treating them as isolated incidents. Complementing these detection-focused models, Sejan et al.~\cite{Sejan2024} focused on optimizing node classification performance in dense \ac{IoT} networks, proposing architectures that enhance the distinguishability of device roles. Moving beyond pure intrusion detection, Shen et al.~\cite{Shen2024} applied \acp{GNN} to the problem of traffic prediction for diverse edge \ac{IoT} data, demonstrating that structural learning captures the spatio-temporal dependencies required for forecasting network load. While these works including Zola et al.~\cite{Zola2022} and Carletti et al.~\cite{Carletti2025} significantly advanced the state of the art in specific sub-tasks, they generally treat the  \ac{GNN} as a black box, outputting verdicts without exposing the underlying structural dependencies.

Current research frontiers are attempting to bridge this interpretability gap through advanced representation learning and \ac{XAI}. Gioacchini et al. have extended their framework to include cross-network embedding transfer~\cite{Gioacchini2024} and generic multi-modal representations~\cite{Gioacchini2024a} to enhance model generalization. Furthermore, efforts to integrate explainability have led to stacking models that combine complementary embeddings with interpretability mechanisms~\cite{Gioacchini2024b}, or the application of post-hoc feature attribution methods like \ac{SHAP} and \ac{LIME} as discussed by Čermák~\cite{Cermak2021}. However, these approaches largely remain external to the learning process and do not exploit the intrinsic geometry of the data and rarely leverage \ac{MnL} to visually project the high-dimensional latent space of \acp{GNN}. This separation limits the ability to provide an intuitive, topological interface for network monitoring. Addressing this limitation requires a methodology that synthesizes structural representation learning with manifold-preserving projection. Therefore, the following section delineates the theoretical foundations of the algorithmic components selected to bridge this gap.
\section{Algorithms and Architectures}
\label{sec:algorithms}

This section delineates the foundational algorithms relevant to this work. In specific, it covers graph representation with \acp{GNN} (Subsection \ref{subsec:gnn_foundations}), dimensionality reduction with \ac{P-UMAP} (Subsection \ref{subsec:umap}), and feature attribution with \ac{SHAP} (Subsection \ref{subsec:shap}). 

\subsection{Graph Neural Architectures}
\label{subsec:gnn_foundations}

\acp{GNN} represent a class of \ac{DL} architectures designed to operate on data structured as graphs \cite{Hamilton2020}. Unlike standard Convolutional Neural Networks that operate on Euclidean grids, \ac{GNN} models leverage the topological structure of non-Euclidean data to learn low-dimensional vector representations or embeddings for nodes, edges, or the entire graph structure.

The fundamental mechanism driving modern \ac{GNN} architectures consists of the neural message passing framework. In this paradigm, the representation of a node is iteratively updated by exchanging and aggregating information from the local neighborhood of the node. Formally, for a graph $\mathcal{G} = (\mathcal{V}, \mathcal{E})$, the embedding update for a node $v$ at layer $k$ is governed by two principal functions, namely, the message aggregation and the state update.

The message aggregation step computes a message vector $\boldsymbol{m}_v^{(l)}$ based on the features of the neighbors and the connecting edges,

\begin{equation}
\boldsymbol{m}_v^{(l)} = \mathcal{A}^{(l)} \left( \left\{ \phi^{(l)} \left( \boldsymbol{h}_v^{(l-1)}, \boldsymbol{h}_u^{(l-1)}, \boldsymbol{e}_{vu} \right) : u \in \mathcal{N}(v) \right\} \right)
\end{equation}
where $\boldsymbol{h}_v^{(l)}$ denotes the embedding vector of node $v$ at layer $l$, and $\boldsymbol{e}_{vu}$ represents the features of the edge connecting $u$ to $v$. The term $\mathcal{N}(v)$ indicates the set of neighbors of node $v$. The function $\phi^{(l)}$ serves as a learnable message function, and $\mathcal{A}^{(l)}$ acts as a differentiable, permutation-invariant aggregation operator. At layer $l=0$, $h_v^{(0)}$ represents the features associated with node $v$. Subsequently, the update step combines the aggregated message with the current representation of the node to generate the embedding for the next layer,
\begin{equation}
\boldsymbol{h}_v^{(l)} = \gamma^{(l)} \left( \boldsymbol{h}_v^{(l-1)}, \boldsymbol{m}_v^{(l)} \right)
\end{equation}
where $\gamma^{(l)}$ is a differentiable update function that produces the new node state.
To capture the network topology, we utilize the \ac{GIN} architecture \cite{Xu2019}, widely recognized as one of the most expressive \ac{GNN} variants due to its ability to model underlying structural dependencies that simpler architectures often miss. Within \ac{GIN}, the update rule for node $v$ at layer $l$ is defined as follows,

\begin{equation}
    \boldsymbol{h}_v^{(l)} = \text{MLP}^{(l)}\left( (1 + \epsilon^{(l)}) \boldsymbol{h}_v^{(l-1)} + \sum_{u \in \mathcal{N}(v)} \boldsymbol{h}_u^{(l-1)} \right)
\end{equation}
where, $\epsilon^{(l)}$ is a learnable parameter that balances the central node features against the aggregation of its neighbors. The term $\text{MLP}^{(l)}$ refers to a \acl{MLP} applied at layer $l$, which serves as a universal approximator to map the aggregated features into the new representation space. By utilizing a sum aggregator coupled with this non-linear transformation, \ac{GIN} ensures that the structural information is preserved throughout the learning process.

However, while \ac{GIN} effectively encodes this structural complexity, the resulting embeddings reside in a high-dimensional latent space $\mathbb{R}^D$ that remains opaque to human interpretation. To visualize these embeddings, it is necessary to map the data to a low-dimensional space($\mathbb{R}^d$, with $d << D$, without distorting the learned topology. Given that \acp{GNN} model highly non-linear relationships, traditional linear projections such as Principal Component Analysis are insufficient. Manifold learning techniques can be suitable to preserve the intrinsic geometry of the data during projection.

\subsection{Parametric Manifold Approximation and Projection}
\label{subsec:umap}

To address this non-linearity, \ac{P-UMAP} is a dimensionality reduction technique grounded in Riemannian geometry and algebraic topology \cite{McInnes2018b,Sainburg2020}. This method constructs a fuzzy topological representation of the high-dimensional data and seeks an equivalent topological structure in the low-dimensional embedding.
While the standard \ac{UMAP} algorithm is transductive and learns a specific set of embeddings for a fixed dataset, streaming applications require the ability to process continuous streams of new network states without computationally prohibitive re-optimization. Therefore, parametric variants are particularly advantageous for such scenarios.

The architecture of \ac{P-UMAP} employs a neural network encoder to learn a parametric mapping $f: \mathbb{R}^D \rightarrow \mathbb{R}^d$. The weights of the network are optimized to minimize the \ac{UMAP} loss function, defined as the cross-entropy between the high-dimensional topological graph and the low-dimensional representation,

\begin{equation}
C = \sum_{i,j} w_{ij} \log\left(\frac{w_{ij}}{v_{ij}}\right) + (1-w_{ij}) \log\left(\frac{1-w_{ij}}{1-v_{ij}}\right)
\end{equation}
here $w_{ij}$ represents the membership strength in the high-dimensional simplex, and $v_{ij}$ indicates the probability of the existence of the edge in the low-dimensional space. This parametric approach provides an inherent inductive capability, allowing the projection of new samples into the visualization space rapidly.

However, while dimensionality reduction effectively reduces the dimensionality of the data, allowing for an easy visualization of the embeddings, it does not inherently elucidate the specific input features driving these projections. To handle these visual insights, it is necessary to decompose the model decisions into understandable components.

\subsection{Shapley Additive Explanations}
\label{subsec:shap}

To ensure transparency in decision-making processes, \ac{SHAP} provides local fidelity to latent representations \cite{Shapley1953,Roth1988}. \ac{SHAP} acts as a unified framework for model interpretation, deriving explanations from cooperative game theory.

\ac{SHAP} assigns an importance value to each feature for a specific prediction. These values, known as Shapley values, represent the average marginal contribution of a feature value across all possible feature coalitions. The fundamental property of \ac{SHAP} is additivity, which is expressed as,

\begin{equation}
\phi_j = \sum_{S \subseteq \mathcal{F} \setminus \{j\}} \frac{|S|! (|\mathcal{F}| - |S| - 1)!}{|\mathcal{F}|!} \left( f(S \cup \{j\}) - f(S) \right)
\end{equation}
where $\mathcal{F}$ represents the set of all input features, $S$ denotes a subset of features excluding $j$, and $f(S)$ indicates the prediction of the model using only the features in subset $S$. The term $f(S \cup \{j\}) - f(S)$ captures the marginal contribution of feature $j$, which is weighted by the probability of the coalition $S$ occurring.

Leveraging these values, \ac{SHAP} establishes an additive feature attribution method. This property ensures that the sum of the feature attributions equals the difference between the model prediction and the base value, denoted as $\Delta \text{pred} = \text{pred} - \text{base}$. The baseline value, $\phi_0$, represents the average prediction estimated over the entire training dataset. It acts as the reference point for the additive attribution,
\begin{equation}
g(\boldsymbol{z}') = \phi_0 + \sum_{j=1}^{M} \phi_j z'_j
\end{equation}

where $g$ represents the explanation model, $M$ is the number of input features (the cardinality of the feature set), $\boldsymbol{z}' \in \{0,1\}^M$ denotes the coalition vector, $\phi_j$ is the Shapley value for feature $j$, and $\phi_0$ is the base value of the model.

Within the context of \ac{GNN} architectures, \ac{SHAP} facilitates the decomposition of the output of the model to identify which specific input feature attributes contributed most significantly to the learned embedding or classification. This capability provides local interpretability, enabling the diagnosis of specific network events. 

The following section describes the specific orchestration of the previously described components to construct the proposed framework.
\section{Dataset Description}
\label{sec:dataset}

This section describes the data source employed to evaluate the performance of the proposed methodology, with the CICIoT2023 benchmark~\cite{Neto2023} as the core training and validation corpus. This high-fidelity dataset was generated within a controlled testbed environment comprising 105 diverse \ac{IoT} devices, ranging from smart cameras and environmental sensors to home automation controllers. The network topology mimics a realistic smart home configuration. Data acquisition was performed using non-intrusive network taps, ensuring the capture of bidirectional traffic flows without the introduction of latency or packet loss.

To assess the robustness of the intrusion detection capabilities, the dataset incorporates 33 distinct attack scenarios organized into seven behavioral macro-categories. These adversarial activities encompass volumetric threats, such as \ac{DDoS} and \ac{DoS} floods, designed to exhaust network bandwidth. Furthermore, the corpus includes network reconnaissance operations (port scanning, vulnerability assessment), access manipulation attempts via brute force and spoofing mechanisms (ARP, DNS), web-based exploits (SQL Injection, XSS), and coordinated Mirai botnet propagation. This stratification enables the evaluation of the model across a wide spectrum of threat vectors, from high-rate flooding to subtle infiltration attempts. Table~\ref{tab:class_distribution} summarizes the distribution of samples per macro-category within the CICIoT2023 dataset.

\begin{table}[ht]
    \centering
    \caption{Distribution of samples per macro-category in the CICIoT2023 dataset.}
    \label{tab:class_distribution}
    \begin{tabular}{l l r}
        \hline
        \textbf{Category} & \textbf{Description} & \textbf{Total Samples} \\
        \hline
        \textbf{DDoS} & Distributed Denial of Service floods & 33,984,560 \\
        \textbf{DoS} & Single-source Denial of Service attacks & 8,090,738 \\
        \textbf{Mirai} & Botnet-specific propagation and flooding & 2,634,124 \\
        \textbf{Benign} & Legitimate user and idle device traffic & 1,098,195 \\
        \textbf{Spoofing} & Identity manipulation (ARP, DNS) & 486,504 \\
        \textbf{Recon} & Network reconnaissance and scanning & 354,565 \\
        \textbf{Web} & Application-layer exploits & 24,829 \\
        \textbf{Brute Force} & Credential exhaustion attacks & 13,064 \\
        \hline
    \end{tabular}
\end{table}

In terms of volume and diversity, the dataset spans approximately 548 GB of raw packet captures collected over a four-week period. This extensive temporal coverage captures diverse protocol interactions (IPv4/IPv6, TCP, UDP, ICMP) and varying usage patterns. The distribution maintains a representative sample between the aforementioned malicious activities and over 16 hours of benign operational traffic, providing a rigorous basis for the training and validation of the anomaly detection algorithms.

Given this diverse operational landscape, the dataset serves as the foundation for the experimental analysis presented in Section~\ref{sec:experiments}, where the proposed methodology and interpretation modules is quantitatively and qualitatively assessed.

\section{Methodology}
\label{sec:methodology}

\begin{figure*}
    \centering
    \includegraphics[width=\linewidth, keepaspectratio]{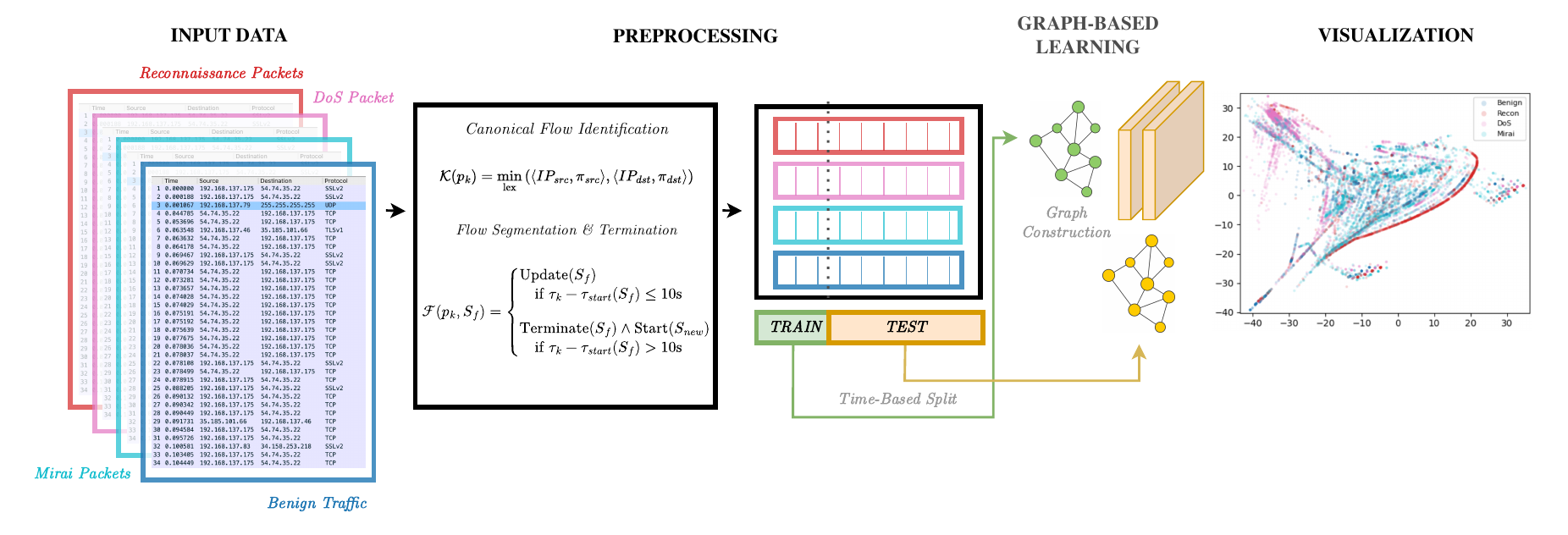}
    \caption{Overview of the proposed pipeline for \ac{IoT} network topology and traffic flow representation. The system ingests raw network traffic and performs canonical flow identification via lexicographical ordering. It then constructs a multigraph representation and learns joint embeddings via a coupled \ac{GNN} and \ac{P-UMAP} architecture, and applies \ac{SHAP} for feature attribution to explain the resulting visual insights.}
    \label{fig:pipeline}
\end{figure*}

In this section, the architectural framework designed to transform raw network traffic into embeddings is described. The approach orchestrates a three-stage pipeline encompassing data ingestion and graph generation, joint embedding learning of devices and traffic flows, and stochastic feature attribution. Figure~\ref{fig:pipeline} illustrates the overall architecture of the integrated pipeline. The architecture consists of three main components

\begin{enumerate}
    \item Data Ingestion and Graph Construction involves the processing of raw network traffic into a structured multigraph representation.
    \item Joint Embedding Learning entails the encoding of device and flow features into a low-dimensional manifold.
    \item Explainability and Feature Attribution provides a feature importance analysis to identify key factors influencing representations.
\end{enumerate}

The subsequent sections detail the mathematical formulation and operational mechanics of these modules. The data acquisition phase is formalized first, where unstructured packet streams are transformed into the multigraph structures that serve as the input for the neural architecture.

\subsubsection{Canonical Flow Identification}
A canonical lexicographical ordering is employed to handle bidirectional traffic. For any packet $p_k$, a unique flow identifier $\mathcal{K}$ is generated to aggregate forward ($A \to B$) and backward ($B \to A$) traffic into a unified entity. The canonical direction is determined by,
\begin{equation}
\mathcal{K}(p_k) = \min_{\text{lex}}\left( \langle IP_{src}, \pi_{src} \rangle, \langle IP_{dst}, \pi_{dst} \rangle \right)
\end{equation}
where $\min_{\text{lex}}$ denotes the minimum value according to a lexicographical comparison, and the subscripts \textit{src} and \textit{dst} refer to the source and destination attributes (IP address and port $\pi$), respectively. This mapping ensures that all packets belonging to the same session are mapped to a single distinct communication event, regardless of the initiation direction. 

\subsubsection{Flow Segmentation}
To transform the continuous packet stream $\mathcal{P}$ into discrete flow instances, a stateful stream processing engine maintains an active flow table $\mathcal{T}$ mapping the canonical flow key $\mathcal{K}$ to a mutable state object $S_f$. For each incoming packet $p_k$ with timestamp $\tau_k$, the engine queries the flow table and if the key is absent, a new state $S_f$ is initialized with $\tau_{start} = \tau_k$; otherwise, the engine updates state statistics such as packet counts, byte volumes, and inter-arrival times. The directional logic evaluates endpoint ordering to attribute each packet correctly to the forward or backward subvectors.

Defining flow boundaries is critical for feature consistency and online processing. The system implements a fixed time window strategy with a maximum flow duration $\delta_{max} = 10$ seconds instead of relying on protocol specific flags or idle timeouts. A flow $f$ is automatically terminated and flushed to the dataset when its duration exceeds this threshold, such that
\begin{equation}
\tau_k - \tau_{start}(S_f) > 10s.
\end{equation}
This approach guarantees periodic updates every 10 seconds regardless of connection longevity. Upon termination, the accumulated state $S_f$ is frozen and transformed into a feature vector. To capture the characteristics of \ac{IoT} traffic, a total of 115 distinct features are extracted across two entity levels. Table~\ref{tab:feature_definitions} provides a taxonomy of these attributes, categorizing them into volume statistics, temporal dynamics, and protocol specific indicators for both device nodes ($F_v=17$) and traffic flow edges ($F_e=98$).

\begin{table*}[ht]
\centering
\caption{Taxonomy of Extracted Features for Device Nodes ($F_v=17$) and Traffic Flow Edges ($F_e=98$)}
\label{tab:feature_definitions}
\resizebox{\textwidth}{!}{%
\begin{tabular}{l l p{0.65\textwidth}}
\hline
\textbf{Entity} & \textbf{Category} & \textbf{Feature Names / Description} \\
\hline
\hline
\multirow{5}{*}{\textbf{Device}} & \multirow{2}{*}{Traffic Volume} & \texttt{bytes\_out}, \texttt{pkts\_out}, \texttt{bytes\_in}, \texttt{pkts\_in} (Total counts) \\
\textbf{($\boldsymbol{X}$)} & & \texttt{mean\_bytes\_out}, \texttt{std\_bytes\_out}, \texttt{mean\_bytes\_in}, \texttt{std\_bytes\_in} (Packet size stats) \\
\cline{2-3}
 & Rate Dynamics & \texttt{mean\_bps\_out}, \texttt{mean\_pps\_out}, \texttt{mean\_bps\_in}, \texttt{mean\_pps\_in} \\
 & Protocol Flags & \texttt{tcp\_flag}, \texttt{udp\_flag}, \texttt{icmp\_flag} (Boolean activity indicators) \\
 & Symmetry Ratios & \texttt{ratio\_bytes\_out\_in}, \texttt{ratio\_flows\_out\_in} \\
\hline
\hline
\multirow{12}{*}{\textbf{Flow}} & Global Stats & \texttt{duration}, \texttt{dur\_zero}, \texttt{pkts}, \texttt{bytes}, \texttt{bytes\_per\_pkt}, \texttt{pps}, \texttt{bps} \\
\textbf{($\boldsymbol{E}$)} & Packet Sizing & \texttt{size\_min}, \texttt{size\_max}, \texttt{size\_mean}, \texttt{size\_var} \\
 & Inter-Arrival Time & \texttt{iat\_mean}, \texttt{iat\_std}, \texttt{iat\_min}, \texttt{iat\_max} \\
 \cline{2-3}
 & \multirow{3}{*}{Directional (A$\to$B)} & \texttt{dir\_a2b\_pkts}, \texttt{dir\_a2b\_bytes}, \texttt{dir\_a2b\_pps}, \texttt{dir\_a2b\_bps} \\
 & & \textit{Sizing:} \texttt{dir\_a2b\_size\_[mean, var, min, max]} \\
 & & \textit{Timing:} \texttt{dir\_a2b\_iat\_[mean, std, min, max]} \\
 \cline{2-3}
 & \multirow{3}{*}{Directional (B$\to$A)} & \texttt{dir\_b2a\_pkts}, \texttt{dir\_b2a\_bytes}, \texttt{dir\_b2a\_pps}, \texttt{dir\_b2a\_bps} \\
 & & \textit{Sizing:} \texttt{dir\_b2a\_size\_[mean, var, min, max]} \\
 & & \textit{Timing:} \texttt{dir\_b2a\_iat\_[mean, std, min, max]} \\
 \cline{2-3}
 & L4 Protocols & One-hot encoded: \texttt{proto\_tcp}, \texttt{proto\_udp}, \texttt{proto\_icmp}, \texttt{proto\_icmpv6} \\
 & \multirow{2}{*}{Port Categories} & One-hot encoded service categories for Source and Destination (e.g., \texttt{http}, \texttt{ssh}, \texttt{dns}, \texttt{mqtts}, \texttt{ftp}, \texttt{dhcp}, \texttt{adb}, \texttt{telnet}). Includes specialized ranges (\texttt{ephemeral}, \texttt{unknown}) and variants (\texttt{alt}, \texttt{alt2}). \\
\hline
\end{tabular}
}
\end{table*}

\subsubsection{Homogeneous Multigraph Construction}
The network snapshot is modeled as a homogeneous  multigraph $\mathcal{G} = (\mathcal{V}, \mathcal{E})$ specifically structured to handle multiple concurrent connections between devices. In this representation, the set of nodes $\mathcal{V} = \{v_1, \dots, v_N\}$ encompasses the unique devices identified by their IP addresses. The set of edges $\mathcal{E} = \{e_1, \dots, e_M\}$ accounts for the communication flows where each edge corresponds to a distinct flow $f_k$ derived from the canonical key $\mathcal{K}$ and finalized by the termination logic. This multigraph formulation allows the model to preserve the structural complexity of the network by representing every individual traffic exchange as a unique connection between device nodes.

Unlike simple graphs, multiple edges are allowed between two nodes $v_i$ and $v_j$ if distinct flows exist (e.g., concurrent SSH and HTTP sessions). The connectivity is defined by the coordinate tensor $\boldsymbol{A} \in \mathbb{N}^{2 \times M}$,
\begin{equation}
\boldsymbol{A} = \begin{bmatrix} 
s_1 & s_2 & \cdots & s_M \\
t_1 & t_2 & \cdots & t_M 
\end{bmatrix}
\end{equation}
where $s_k$ and $t_k$ are the indices of the source and target devices for the $k$-th flow.

\subsubsection{Feature Matrix Representation}
The data are structured into two primary tensors corresponding to the node attributes and edge attributes, where vectors are composed by concatenating distinct functional groups.

\paragraph{Device (Node) Matrix}
Let $N$ be the number of unique devices. For each device $v_i$, the feature vector $\boldsymbol{x}_i \in \mathbb{R}^{17}$ is constructed by aggregating traffic statistics. The vector is decomposed into three semantic groups: traffic volume \& rates (12 dims), protocol activity (3 dims), and symmetry ratios (2 dims);

\begin{equation}
\boldsymbol{x}_i = \Big[ 
    \underbrace{\Psi_{vol}(Out, In) \parallel \Psi_{rate}}_{\text{Volume \& Rate Stats}} \parallel 
    \underbrace{\boldsymbol{f}_{proto}}_{\text{Flags}} \parallel 
    \underbrace{\boldsymbol{r}_{sym}}_{\text{Ratios}} 
\Big]
\end{equation}
where $\boldsymbol{f}_{proto}$ contains boolean indicators for TCP/UDP/ICMP, and $\boldsymbol{r}_{sym}$ captures behavioral symmetry (e.g., $\frac{bytes_{out}}{bytes_{in}}$). The global node matrix is defined as,
\begin{equation}
\boldsymbol{X} = \begin{bmatrix} 
\boldsymbol{x}_1 \\ 
\vdots \\ 
\boldsymbol{x}_N 
\end{bmatrix} \in \mathbb{R}^{N \times 17}
\end{equation}

\paragraph{Flow (Edge Attribute) Matrix }
Each edge $e_k$ carries a high-dimensional feature vector $\boldsymbol{e}_k \in \mathbb{R}^{98}$. To capture the dynamics of the session, this vector is formed by concatenating global session counters, fine-grained bidirectional statistics, and categorical encodings.

\begin{equation}
\boldsymbol{e}_k = \Big[ 
    \underbrace{\boldsymbol{s}_{glob}}_{\text{Global}} \parallel 
    \underbrace{\boldsymbol{s}_{A \to B} \parallel \boldsymbol{s}_{B \to A}}_{\text{Bi-Directional Dynamics}} \parallel 
    \underbrace{\Gamma(\rho) \parallel \Gamma(\pi)}_{\text{Proto \& Ports}} 
\Big]
\end{equation}
where $\boldsymbol{s}_{A \to B}$ and $\boldsymbol{s}_{B \to A}$ include specific IAT and packet size distributions for each direction, and $\Gamma(\cdot)$ represents the one-hot embedding of protocols and service ports. The resulting edge attribute matrix is
\begin{equation}
\boldsymbol{E} = \begin{bmatrix} 
\boldsymbol{e}_1 \\ 
\vdots \\ 
\boldsymbol{e}_M 
\end{bmatrix} \in \mathbb{R}^{M \times 98}
\end{equation}
This structured formulation allows the GNN to distinguish between structural device properties (node features) and transient communication behaviors (edge features) during message passing.

To prevent temporal information leakage and ensure generalization in real-world streaming environments, the model follows a time-stratified evaluation protocol. The data is sorted chronologically and divided into four sequential partitions of equal size, where the first partition serves as the training set and the subsequent three segments---designated as Test A, Test B, and Test C ---are used as test sets. This temporal split is strictly governed by the condition that every training instance must precede any test instance in time, as expressed by,
\begin{equation}
\begin{split}
    \forall f_{train} \in \mathcal{F}_{train}, \: \forall f_{test} \in \mathcal{F}_{test} \\ 
    \implies \tau_{start}(f_{train}) < \tau_{start}(f_{test})
\end{split}
\end{equation}
Such a strategy preserves the natural evolution of network behavior while simulating a realistic deployment scenario. Following this partitioning, the selected features are fed into the neural architecture to begin the joint representation learning process.

\subsection{Joint Embedding Learning}
\label{subsec:encoder}

\begin{figure*}
    \centering
    \includegraphics[width=\linewidth, keepaspectratio]{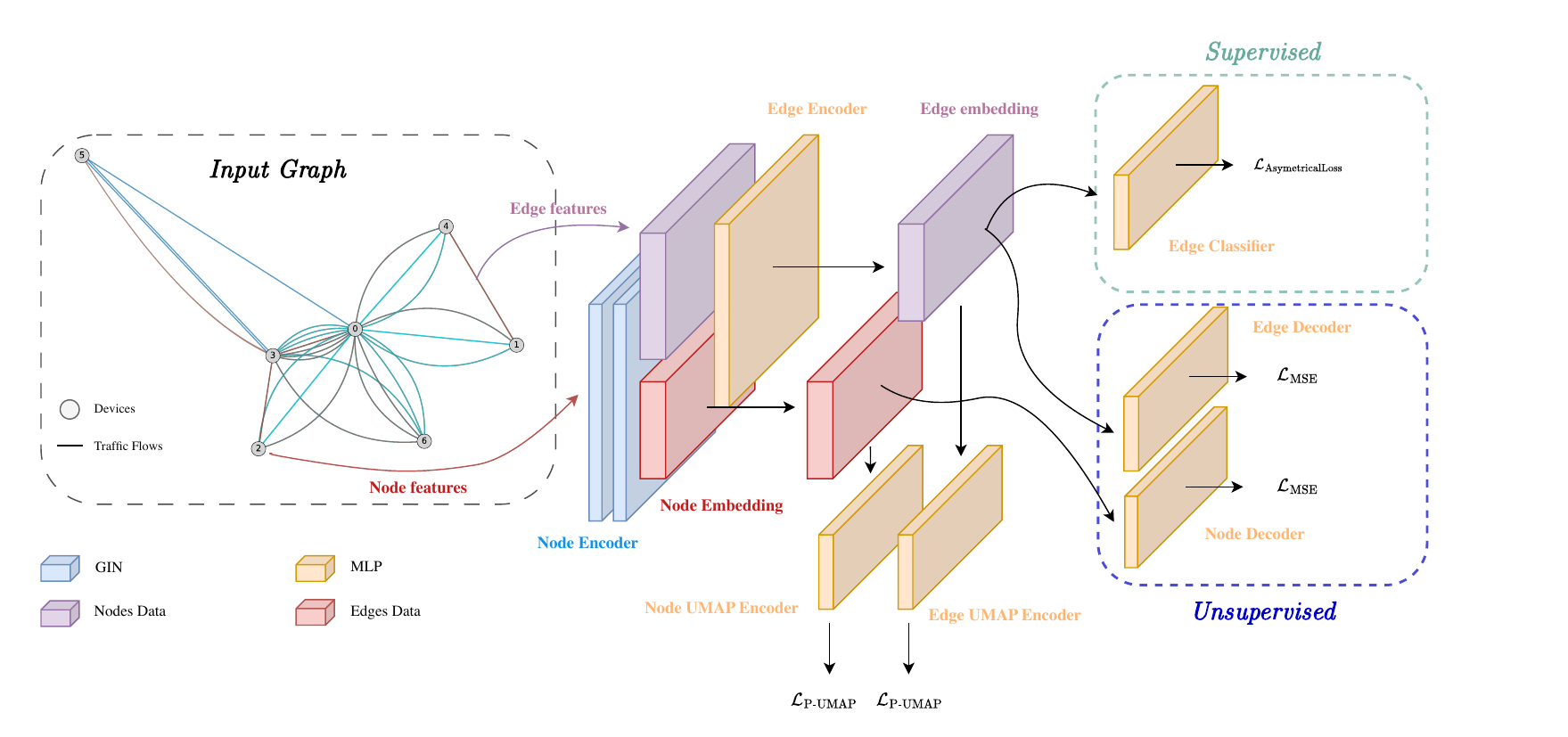}
    \caption{Schematic of the proposed architecture integrating coupled \ac{GIN} and \ac{P-UMAP} for joint device and flow embedding. The model encodes node and edge features, enforces topological consistency, and reconstructs original attributes via dual decoders in an unsupervised setting, or classifies edges minimizing an asymmetric loss in a supervised setting.}
    \label{fig:architecture}
\end{figure*}

The primary objective of the proposed architecture is the simultaneous learning of task-specific predictions and the computation of low dimensional embeddings. Illustrated in Figure ~\ref{fig:architecture}, the neural framework utilizes a \ac{GIN} backbone combined with a \ac{P-UMAP} objective to ensure that the latent space maintains a structure suitable for visualization. To represent the dual nature of network data, the architecture processes device-level features and flow-level attributes through distinct streams to maintain the specificity of each data type. The training process evaluates two distinct variants based on supervised and unsupervised loss functions. The proposed framework ensures that the learned embeddings facilitate the extraction of visual and numerical interpretability insights. In this setting, let $\mathcal{G} = (\mathcal{V}, \mathcal{E})$ be the graph defined previously, with node features $\boldsymbol{X}$ and edge features $\boldsymbol{E}$.

The node encoder utilizes \ac{GIN} operators to capture local structural isomorphism. For a node $v$ at layer $l$, the latent representation $\boldsymbol{h}_v^{(l)}$ is updated iteratively by aggregating neighbor features,
\begin{equation}
    \boldsymbol{h}_v^{(l)} = MLP^{(l)} \left( (1 + \epsilon^{(l)}) \cdot \boldsymbol{h}_v^{(l-1)} + \sum_{u \in \mathcal{N}(v)} \boldsymbol{h}_u^{(l-1)} \right)
\end{equation}
where $\epsilon$ is a learnable scalar and $\mathcal{N}(v)$ denotes the neighborhood of $v$. This produces a final node embedding matrix $\boldsymbol{H} \in \mathbb{R}^{N \times d}$.

To model communication dynamics, a distinct latent representation $\boldsymbol{z}_{uv}$ is computed for every edge $(u, v)$. This is achieved via a fusion function that concatenates the learned embeddings of the incident nodes with the raw edge attributes,
\begin{equation}
    \boldsymbol{z}_{uv} = {MLP}_{\text{edge}} \left( \boldsymbol{h}_u + \boldsymbol{h}_v \parallel \boldsymbol{e}_{uv} \right)
\end{equation}
This results in an edge embedding matrix $\boldsymbol{Z} \in \mathbb{R}^{M \times d'}$, ensuring that every flow is represented by both its specific attributes and the context of its endpoints.

To bridge the gap between high-dimensional feature learning and interpretable visualization, independent parametric projection mechanisms are employed for nodes and edges. The architecture includes dedicated projection heads that map the latent embeddings into low-dimensional manifolds optimized to preserve topological structure.

Let $\psi_{\text{node}}$ and $\psi_{\text{edge}}$ denote two distinct \acp{MLP}. The projected coordinate vectors for nodes ($\boldsymbol{u}_v$) and edges ($\boldsymbol{w}_{uv}$) are computed as,
\begin{equation}
\begin{aligned}
    \boldsymbol{u}_v &= \psi_{\text{node}}(\boldsymbol{h}_v) \in \mathbb{R}^{d_{low}} \\
    \boldsymbol{w}_{uv} &= \psi_{\text{edge}}(\boldsymbol{z}_{uv}) \in \mathbb{R}^{d_{low}}
\end{aligned}
\end{equation}
To enforce structural consistency, the divergence between the graph topology and the projected geometry is minimized separately for each domain. The probability of adjacency between two entities $i$ and $j$ in the projected space is modeled as a differentiable function of their Euclidean distance $d_{ij}$, 
\begin{equation}
    p_{ij} = \left( 1 + a \cdot d_{ij}^{2b} \right)^{-1}
\end{equation}
where $a$ and $b$ are hyperparameters fitted to the initial data distribution. The topological loss $\mathcal{L}_{\text{topo}}$ is formulated as the fuzzy binary cross-entropy between the observed adjacency and the latent probabilities,
\begin{equation}
    \mathcal{L}_{\text{topo}}(\boldsymbol{Y}, \boldsymbol{P}) = \sum_{(i,j) \in \mathcal{Y}^+} \log\left(\frac{1}{p_{ij}}\right) + \sum_{(i,k) \in \mathcal{Y}^-} \log\left(\frac{1}{1 - p_{ik}}\right)
\end{equation}
where $\mathcal{Y}^+$ and $\mathcal{Y}^-$ represent the sets of positive (connected) and negative (disconnected) pairs, respectively. This loss is applied to both the node projections $\boldsymbol{U}$ and the edge projections $\boldsymbol{W}$.

Finally, the architecture is trained under two distinct operational paradigms depending on the availability of ground-truth labels. We formulate a primary task loss $\mathcal{L}_{\text{task}}$ that drives the learning of the latent space, which takes one of two mutually exclusive forms.

In the unsupervised learning variant referred as \ac{GNN-AE}, the model operates as a dual-stream autoencoder aiming to reconstruct the original input attributes. We employ decoders $D_{\text{node}}(\boldsymbol{H}) \to \hat{\boldsymbol{X}}$ and $D_{\text{edge}}(\boldsymbol{Z}) \to \hat{\boldsymbol{E}}$ and minimize the reconstruction error. Assuming a Gaussian noise distribution for the continuous feature space, the loss is defined as the Mean Squared Error (MSE) over both entities,
\begin{equation}
    \mathcal{L}_{\text{MSE}} = \frac{1}{N} \sum_{i=1}^{N} \|\boldsymbol{x}_i - \hat{\boldsymbol{x}}_i\|_2^2 + \frac{1}{M} \sum_{j=1}^{M} \|\boldsymbol{e}_j - \hat{\boldsymbol{e}}_j\|_2^2
\end{equation}

Alternatively, in the supervised learning variant refereed as \ac{GNN-CLS}, when attack labels are available, the reconstruction objective is replaced
by an objective tailored to the specific classification task. The extreme class imbalance typical of intrusion detection data is tackled by using the asymmetric loss \cite{Ridnik2021}. This loss function decouples the focusing mechanism for positive and negative samples, allowing for finer control over the gradient contribution from easy negatives and it is given by
\begin{equation}
\begin{split}
    \mathcal{L}_{\text{Asym}} = - \frac{1}{M} \sum_{i=1}^{M} \sum_{k=1}^{C} \Big[ & y_{i,k} (1 - p_{i,k})^{\gamma_+} \log(p_{i,k}) \\
    & + (1 - y_{i,k}) p_{i,k}^{\gamma_-} \log(1 - p_{i,k}) \Big]
\end{split}
\end{equation}
where $p_{i,k}$ is the predicted probability for class $k$, $y_{i,k}$ is the one-hot encoded target, and $\gamma_+$ and $\gamma_-$ are the focusing parameters for positive and negative samples, respectively.

Regardless of the chosen variant, the primary task loss is jointly optimized with the topological regularization terms to ensure the learned manifold preserves structural coherence.The final objective function is defined as a linear combination of the task-specific loss and the topological regularization terms,
\begin{equation}
    \mathcal{L}_{\text{total}} = \lambda_{\text{task}} \cdot \mathcal{L}_{\text{task}} + \lambda_{\text{topo}} \cdot \left( \mathcal{L}_{\text{topo}}(\boldsymbol{U}) + \mathcal{L}_{\text{topo}}(\boldsymbol{W}) \right)
\end{equation}
where $\mathcal{L}_{\text{task}}$ corresponds to either $\mathcal{L}_{\text{MSE}}$ or $\mathcal{L}_{\text{Asym}}$. The hyperparameters $\lambda_{\text{task}}$ and $\lambda_{\text{topo}}$ weight the contribution of each term.

In summary, by minimizing the composite objective $\mathcal{L}_{\text{total}}$, the proposed architecture learns a unified manifold where geometric proximity encodes both the structural connectivity of devices and the behavioral semantics of traffic flows. The following subsection addresses this challenge by introducing a stochastic feature attribution mechanism adapted for coordinate-based manifold explanations.

\subsection{Explainability and Feature Attribution}
\label{subsec:shap_impl}

To address the interpretability of the learned representations, an interpretability method based on \ac{SHAP} is implemented. Unlike standard classification tasks where \ac{SHAP} explains the probability of a class, here \ac{SHAP} is adapted to explain the coordinates of the embedding in the low-dimensional manifold.

Let $f: \mathbb{R}^F \to \mathbb{R}^d$ be the composite function of the \ac{GNN} encoder and the parametric \ac{UMAP} projector, mapping an input feature vector $\boldsymbol{x}$ to an embedding $\boldsymbol{z}$. The goal is to quantify the contribution $\phi_{ij}$ of each input feature $j$ to each output dimension $i$ of the embedding.

A Monte Carlo approximation strategy is adopted to estimate these Shapley values \cite{Strumbelj2014}. For a specific flow or device embedding, the marginal contribution of feature $j$ to dimension $i$ is computed by simulating feature coalitions. Binary masks $\boldsymbol{m} \in \{0,1\}^F$ are sampled and the input features are perturbed using a background distribution $\mathcal{D}_{bg}$ derived from the training data,
\begin{equation}
\phi_{ij}(\boldsymbol{x}) \approx \frac{1}{K} \sum_{k=1}^{K} \left[ f_i(\boldsymbol{x} \odot \boldsymbol{m}_k^{(+j)}) - f_i(\boldsymbol{x} \odot \boldsymbol{m}_k^{(-j)}) \right]
\end{equation}
where $f_i$ denotes the $i$-th component of the mapping function $f$, and $K$ is the number of Monte Carlo samples and $m_x^{(+j)}$ and $m_x^{(-j)}$ represent respectively the mask with $(+j)$ and without $(-j)$ the $j^{th}$ feature. This stochastic process enables the quantification of the directional influence each network feature exerts on the position of a node in the \ac{UMAP} visualization.

With the theoretical framework fully established, covering data ingestion, joint embedding learning, and feature attribution. The following section introduces the realistic dataset selected for this evaluation.
\section{Experiments and Results}
\label{sec:experiments}

This section presents the experimental evaluation of the proposed Graph Neural Network pipeline. The assessment follows a quantitative to qualitative workflow that begins with measuring the structural quality of the embeddings to validate the representation learning capabilities. This is followed by a granular analysis of classification performance and topological evolution. Finally, an interpretability analysis is conducted on the learned manifold.

\subsection{Quantitative Assessment of Latent Space}

To validate the structural coherence of the generated embeddings, the proposed architectures \ac{GNN-AE} and \ac{GNN-CLS} were compared against a baseline approach where \ac{P-UMAP} is applied directly to the raw feature space. Two clustering validity indices were computed: the \ac{DBI}, where lower values indicate better separation, and the Silhouette Score, where higher values indicate better cohesion.

Table~\ref{tab:embedding_metrics} summarizes the results for both traffic flows (Edges) and IoT devices (Nodes). The data indicate that graph-based embeddings yield lower \ac{DBI} and higher Silhouette scores compared to the raw data baseline across all metrics. For Traffic Flows (Edges), the supervised \ac{GNN-CLS} model presents the lowest \ac{DBI} ($4.227 \pm 0.652$) and highest Silhouette ($0.142 \pm 0.068$), suggesting that the classification objective enforces separability between benign and malicious traffic in the latent space.

\begin{table*}[ht]
\centering
\caption{Comparison of \ac{DBI} and Silhouette Score for Edge and Node entities across temporal partitions.}
\label{tab:embedding_metrics}
\begin{tabular}{l c c c c}
\toprule
& \multicolumn{2}{c}{\textbf{Traffic Flows (Edges)}} & \multicolumn{2}{c}{\textbf{IoT Devices (Nodes)}} \\
\cmidrule(lr){2-3} \cmidrule(lr){4-5}
\textbf{Model} & \textbf{\ac{DBI}} $\downarrow$ & \textbf{Silhouette} $\uparrow$ & \textbf{\ac{DBI}} $\downarrow$ & \textbf{Silhouette} $\uparrow$ \\
\midrule
P-UMAP (Baseline) & 20.249 $\pm$ 10.534 & 0.032 $\pm$ 0.074 & 1.331 $\pm$ 0.265 & 0.085 $\pm$ 0.025 \\
\ac{GNN-AE} (Unsupervised) & 6.325 $\pm$ 3.657 & 0.021 $\pm$ 0.045 & \textbf{1.553 $\pm$ 0.076} & \textbf{0.955 $\pm$ 0.005}\\
\ac{GNN-CLS} (Supervised) & \textbf{4.227 $\pm$ 0.652} & \textbf{0.142 $\pm$ 0.068} & 2.117 $\pm$ 0.291 & 0.665 $\pm$ 0.120 \\
\bottomrule
\end{tabular}%
\end{table*}

This quantitative superiority in clustering metrics suggests that the embeddings learned by the supervised model are not only topologically consistent but also optimized for decision boundaries. To verify this, the following subsection investigates how this structural quality translates into classification performance.

\subsection{Classification Performance and Topology Analysis}

The predictive capabilities of the \ac{GNN-CLS} model were evaluated to assess global efficacy, topological coherence, and robustness against evolving traffic patterns.

\subsubsection{Classification Performance and Latent Structure}
Table~\ref{tab:classification_metrics} summarizes the aggregated performance. A distinct divergence is observed between detection capabilities (Binary) and categorization granularity (Multiclass). The model achieves a mean Binary F1-Score of 0.830, confirming its reliability as a first-line intrusion detection system. However, the multiclass attribution suffers from high variance (Macro F1 Std Dev: $\pm$ 0.093).

\begin{figure*}[ht]
    \centering
    \includegraphics[width=\textwidth]{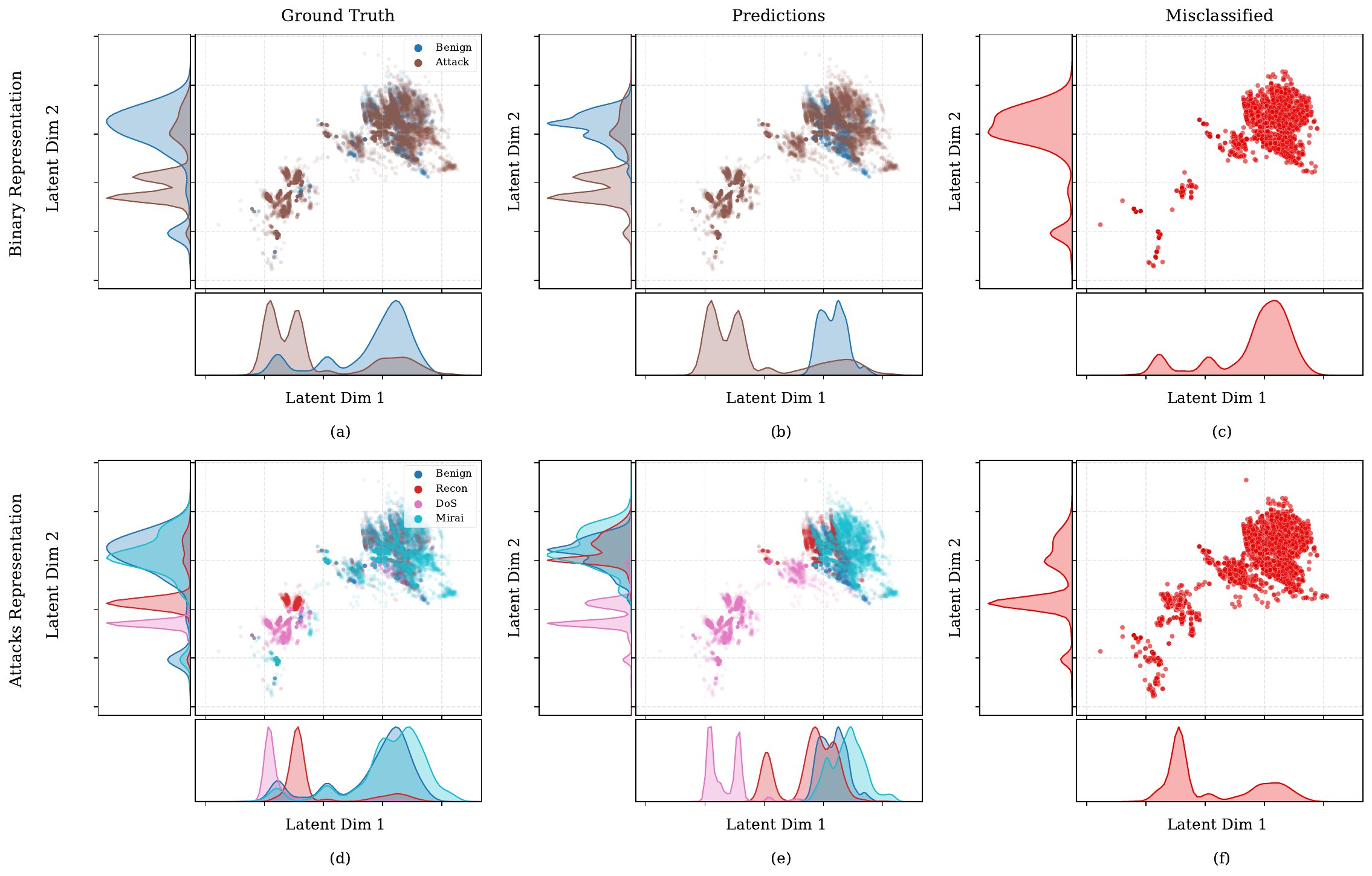}
    \caption{Comparison of Ground Truth, Model Predictions, and Misclassification distribution. Top Row (Binary): Demonstrates clear separability between Benign and Attack traffic, with minimal errors (c). Bottom Row (Multiclass): Reveals structural overlaps between attack classes (e.g., Mirai and DoS), where errors are densely concentrated (f), indicating semantic ambiguity rather than model failure.}
    \label{fig:topology_analysis}
\end{figure*}

To explain this divergence, Figure~\ref{fig:topology_analysis} illustrates the latent topology of the test set by contrasting ground truth labels against model predictions and error distribution. The top row of the visualization demonstrates binary robustness where panels (a) and (b) show an approximately linear separability between benign and attack manifolds. Marginal density plots confirm that these distributions are relatively non-overlaping and result in a nearly error free decision boundary even if several examples are misclassified (c) due to the partial superposition of benign and attack points also in the ground truth projected representations (e). In contrast the bottom row reveals significant multiclass ambiguity as panels (d) and (e) exhibit structural overlaps within the attack categories. While benign traffic remains distinct the reconnaissance and mirai clusters show strong cohesion with denial of service regions. Panel (f) further highlights that misclassifications are not randomly distributed but instead appear structurally concentrated at the boundaries of these overlapping clusters which explains the observed variance in multiclass performance metrics.

\begin{table}[ht]
\centering
\caption{Classification Performance Comparison of Binary Detection and Multiclass Categorization F1-Scores (Mean $\pm$ Std).}
\label{tab:classification_metrics}
\begin{tabular}{lcc}
\toprule
\textbf{Metric} & \textbf{Mean Score} & \textbf{Std Dev} \\
\midrule
\multicolumn{3}{c}{\textit{Binary Detection}} \\
Binary F1 & 0.830 & $\pm$ 0.057 \\
Weighted F1 & 0.767 & $\pm$ 0.050 \\
\midrule
\multicolumn{3}{c}{\textit{Multiclass Categorization}} \\
Macro F1 & 0.564 & $\pm$ 0.093 \\
\addlinespace
\multicolumn{3}{l}{\textit{Per-Class F1 (Macro)}} \\
\hspace{3mm} Benign & 0.469 & $\pm$ 0.039 \\
\hspace{3mm} Recon & 0.289 & $\pm$ 0.150 \\
\hspace{3mm} DoS & 0.845 & $\pm$ 0.133 \\
\hspace{3mm} Mirai & 0.652 & $\pm$ 0.052 \\
\bottomrule
\end{tabular}
\end{table}

While the global metrics indicate satisfactory binary detection, the variance in multiclass performance requires a deeper investigation. Given that network traffic is dynamic, this variability likely stems from temporal shifts in attack behavior rather than static model deficiencies.

\subsubsection{Temporal Evolution and Concept Drift}

\begin{figure*}[ht]
    \centering
    \includegraphics[width=\textwidth]{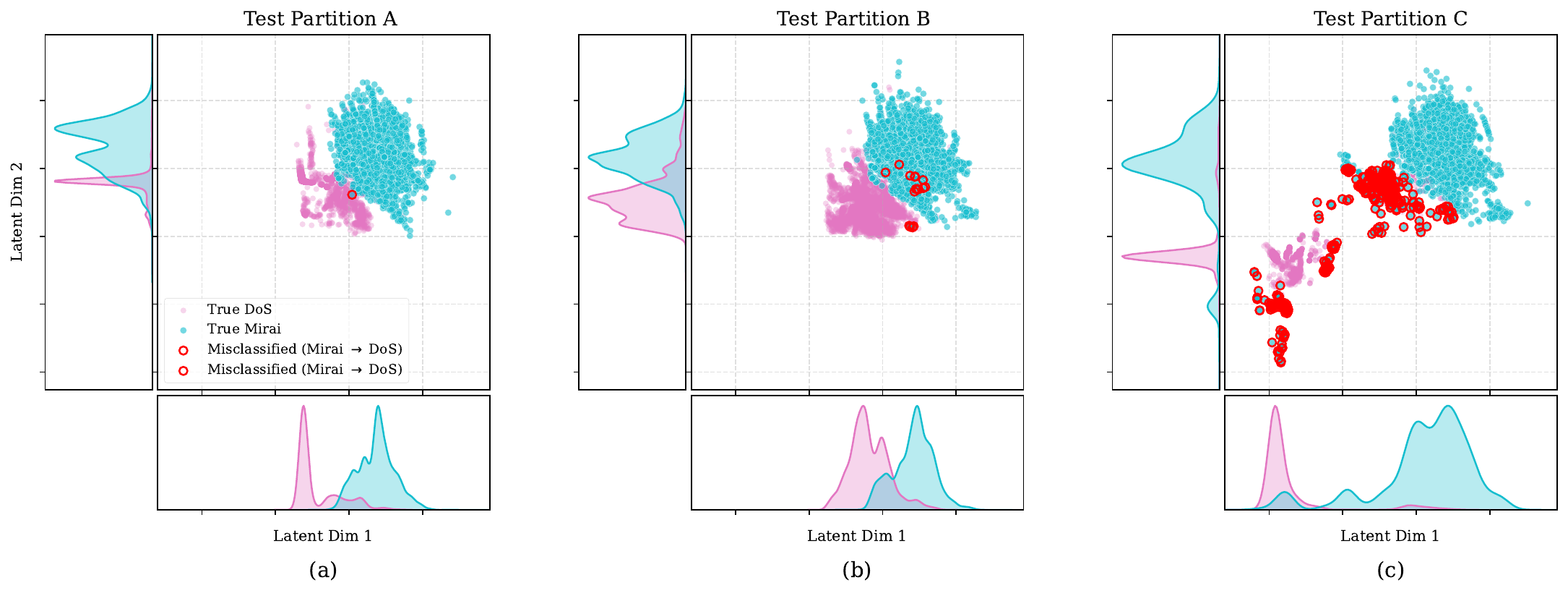}
    \caption{Evolution of latent embeddings across three temporal partitions (Mirai vs Dos) reveals the mechanism of performance degradation. (a) Initially, Mirai and DoS are topologically distinct. (b) The Mirai cluster begins to migrate towards the DoS region. (c) In the final stage, Mirai exhibits mimetic behavior, structurally overlapping with DoS. Red markers highlight misclassified instances, demonstrating that model errors are non-random and localized in the intersection zone between Mirai and DoS.}
    \label{fig:concept_drift}
\end{figure*}

Table~\ref{tab:drift_metrics} breaks down performance across the three chronological partitions (A, B, C) to investigate the source of variance. The results reveal an inverse drift phenomenon: binary performance improves over time (0.773 to 0.908), while Reconnaissance classification collapses (0.458 to 0.093). This degradation is visually corroborated in Figure~\ref{fig:concept_drift} which maps the topological evolution of the latent space across the sequential evaluation segments. Panel (a) presents the results corresponding to test partition A where the Mirai category forms a distinct cluster that is topologically orthogonal to generic denial of service traffic allowing the model to successfully distinguish both threat types. This spatial separation begins to dissolve in panel (b) which represents test partition B and captures the onset of the drift as the Mirai manifold starts to migrate towards the denial of service region. By the final stage illustrated in panel (c) for test partition C, the traffic undergoes a complete topological shift and exhibits mimetic behavior by overlapping with the denial of service cluster. The accompanying error analysis confirms that misclassifications are strictly localized within this intersection zone which validates that the performance drop is driven by the structural unification of attack signatures through volumetric flooding rather than a failure in the underlying model architecture.

The observation of this topological drift raises a critical question: what specific features drive the migration of the Mirai cluster towards the DoS region. To answer this, we move from topological observation to semantic explanation using the proposed interpretability module.

\begin{table}[ht]
\centering
\caption{Evolution of F1-Scores across temporal partitions (A, B, C) illustrating Concept Drift impact.}
\label{tab:drift_metrics}
\begin{tabular}{l c c c}
\toprule
& \multicolumn{3}{c}{\textbf{Temporal Partition (F1-Score)}} \\
\textbf{Class / Task} & \textbf{Part A (Start)} & \textbf{Part B (Mid)} & \textbf{Part C (End)} \\
\midrule
\multicolumn{4}{l}{\textit{Binary Detection Task}} \\
Benign vs. Attack & 0.773 & 0.810 & \textbf{0.908} \\
\midrule
\multicolumn{4}{l}{\textit{Multiclass Attribution Task}} \\
Benign & \textbf{0.516} & 0.471 & 0.421 \\
Recon & \textbf{0.458} & 0.314 & 0.093 \\
DoS & \textbf{0.969} & 0.906 & 0.661 \\
Mirai & \textbf{0.715} & 0.654 & 0.588 \\
\bottomrule
\end{tabular}%
\end{table}

\subsection{Interpretable Embedding Visualization}

To analyze the semantic causes of the topological overlaps observed in Figures~\ref{fig:topology_analysis} and~\ref{fig:concept_drift}, an interpretability analysis was conducted. Figure~\ref{fig:shap_dashboard} presents a comprehensive dashboard that integrates the UMAP latent projection with SHAP feature attribution.

\subsubsection{Semantic Decoding of Latent Axes}
The visualization delineates high density regions for threat categories using 90\% mass isocontours. These latent axes are semantically decoded through a global feature importance analysis that identifies the primary drivers of the embedding space. The horizontal axis acts as a protocol discriminator where positive values are driven by domain name system amplification artifacts and negative values correlate with management protocols. Simultaneously the vertical axis captures traffic complexity by separating internet of things control traffic driven by encrypted message queuing telemetry transport from volumetric flooding characterized by packet rate. The specific features driving these projections are detailed in Tables~\ref{tab:umap_drivers_x} and~\ref{tab:umap_drivers_y}, which isolate the top contributors for each quadrant.

\begin{table*}[ht]
\centering
\caption{Analysis of UMAP Dimension 1 Drivers (Horizontal Axis). Features pushing embeddings towards the Left ($\leftarrow$) or Right ($\rightarrow$) based on SHAP importance.}
\label{tab:umap_drivers_x}
\resizebox{\textwidth}{!}{%
\begin{tabular}{l l l l l}
\toprule
\textbf{Class} & \textbf{Top 1 Driver ($X_1$)} & \textbf{Top 2 Driver ($X_2$)} & \textbf{Top 3 Driver ($X_3$)} & \textbf{Top 4 Driver ($X_4$)} \\
\midrule
\textbf{Benign} & \texttt{dur\_zero} (-0.19) $\leftarrow$ & \texttt{src\_port\_cat\_smtp} (-0.19) $\leftarrow$ & \texttt{dir\_a2b\_iat\_std} (0.18) $\rightarrow$ & \texttt{proto\_tcp} (0.16) $\rightarrow$ \\
\addlinespace
\textbf{Mirai} & \texttt{src\_port\_cat\_smtp\_alt} (-0.22) $\leftarrow$ & \texttt{dir\_a2b\_size\_var} (-0.21) $\leftarrow$ & \texttt{src\_port\_cat\_snmp} (-0.21) $\leftarrow$ & \texttt{dur\_zero} (-0.19) $\leftarrow$ \\
\addlinespace
\textbf{DoS} & \texttt{dst\_port\_cat\_dns} (0.32) $\rightarrow$ & \texttt{pkts} (-0.28) $\leftarrow$ & \texttt{bytes} (0.20) $\rightarrow$ & \texttt{src\_port\_cat\_smtp\_alt} (-0.17) $\leftarrow$ \\
\addlinespace
\textbf{Scan} & \texttt{src\_port\_cat\_smtp\_alt} (-0.35) $\leftarrow$ & \texttt{dir\_a2b\_size\_var} (-0.21) $\leftarrow$ & \texttt{dst\_port\_cat\_irc\_alt} (0.20) $\rightarrow$ & \texttt{src\_port\_cat\_snmp\_trap} (-0.19) $\leftarrow$ \\
\bottomrule
\end{tabular}%
}
\end{table*}

\begin{table*}[ht]
\centering
\caption{Analysis of UMAP Dimension 2 Drivers (Vertical Axis). Features pushing embeddings Upwards ($\uparrow$) or Downwards ($\downarrow$) based on SHAP importance.}
\label{tab:umap_drivers_y}
\resizebox{\textwidth}{!}{%
\begin{tabular}{l l l l l}
\toprule
\textbf{Class} & \textbf{Top 1 Driver ($Y_1$)} & \textbf{Top 2 Driver ($Y_2$)} & \textbf{Top 3 Driver ($Y_3$)} & \textbf{Top 4 Driver ($Y_4$)} \\
\midrule
\textbf{Benign} & \texttt{dur\_zero} (0.30) $\uparrow$ & \texttt{src\_port\_cat\_smtp\_alt} (0.24) $\uparrow$ & \texttt{duration} (0.22) $\uparrow$ & \texttt{src\_port\_cat\_irc\_alt2} (0.21) $\uparrow$ \\
\addlinespace
\textbf{Mirai} & \texttt{src\_port\_cat\_smtp\_alt} (0.29) $\uparrow$ & \texttt{src\_port\_cat\_snmp} (0.25) $\uparrow$ & \texttt{dur\_zero} (0.19) $\uparrow$ & \texttt{src\_port\_cat\_mqtts} (0.17) $\uparrow$ \\
\addlinespace
\textbf{DoS} & \texttt{src\_port\_cat\_mqtts} (0.28) $\uparrow$ & \texttt{size\_min} (-0.20) $\downarrow$ & \texttt{bps} (-0.19) $\downarrow$ & \texttt{dir\_a2b\_iat\_std} (-0.19) $\downarrow$ \\
\addlinespace
\textbf{Scan} & \texttt{dir\_a2b\_size\_var} (0.30) $\uparrow$ & \texttt{src\_port\_cat\_smtp\_alt} (0.28) $\uparrow$ & \texttt{dur\_zero} (0.27) $\uparrow$ & \texttt{src\_port\_cat\_irc\_alt2} (0.24) $\uparrow$ \\
\bottomrule
\end{tabular}%
}
\end{table*}

\subsubsection{Visual Explanation of Behavioral Overlaps}
\begin{figure*}[ht]
    \centering
    \includegraphics[width=\linewidth]{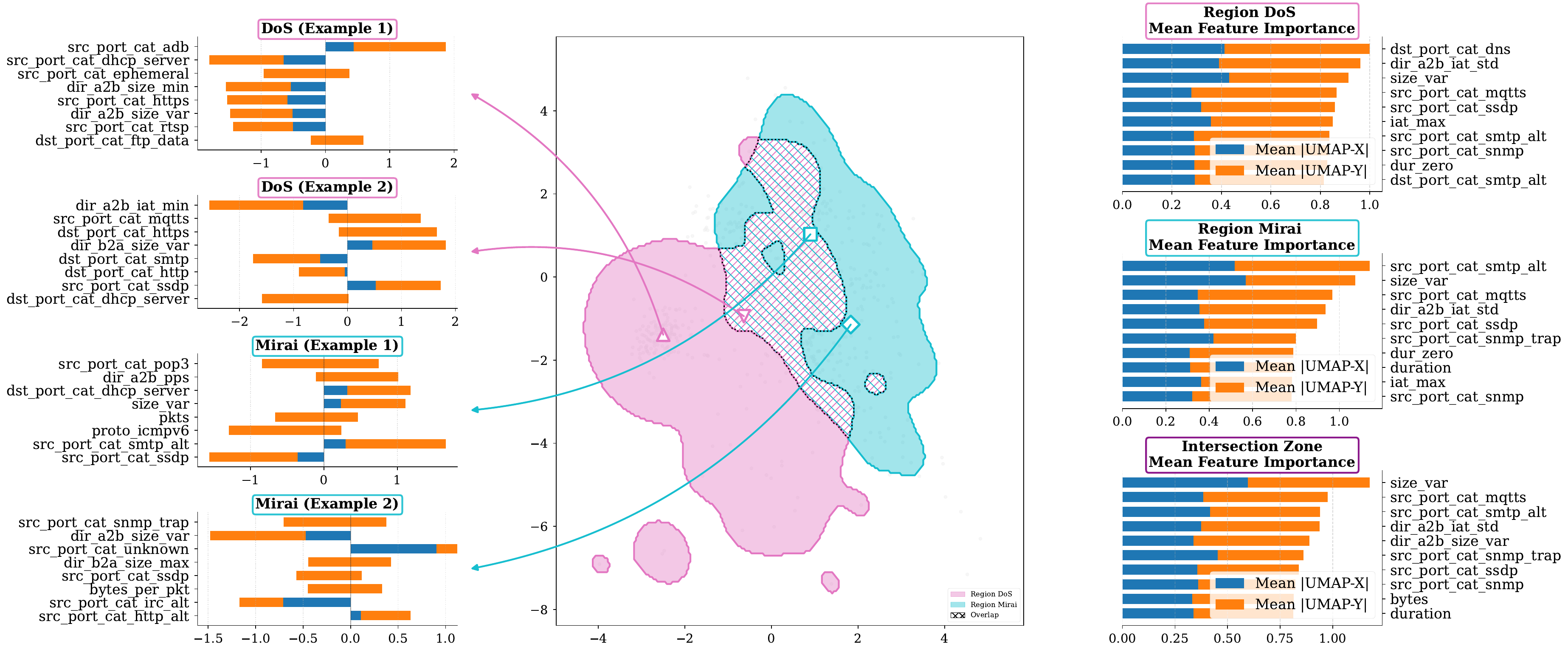}
    \caption{Multi-level interpretability dashboard of the learned GNN latent space. (Center) UMAP projection delineating density regions for DoS (Pink) and Mirai (Cyan). (Right) Global feature importance ranking for pure and intersection zones. (Left) Local SHAP feature attribution.}
    \label{fig:shap_dashboard}
\end{figure*}

The dashboard highlights a critical Intersection Zone (hatched area in Figure~\ref{fig:shap_dashboard}, Center) between DoS (Pink) and Mirai (Cyan). The global feature importance analysis (Right Panel) reveals that this area is characterized by shared features such as packet size variance (\texttt{size\_var}) and protocol patterns (\texttt{mqtts}). This confirms that the mimetic behavior identified in the concept drift analysis is caused by the convergence of botnet attack vectors towards generic UDP/TCP flooding characteristics, creating an irreducible error floor for the multiclass classifier in the absence of deeper packet inspection.

In summary, the experimental analysis validates the proposed pipeline on multiple fronts. Quantitatively, the learned embeddings exhibit high structural coherence and support binary detection. Qualitatively, the topological analysis reveals that performance drops are not random failures but consequences of demonstrable concept drift (mimetic behavior).
\section{Discussion and Conclusion}
\label{sec:discussion}
The experimental results detailed in the previous section underscore the dual capability of the proposed \ac{GNN} architecture: providing robust intrusion detection while offering granular interpretability of the evolving network topology. Building on this validation, this concluding section synthesizes the key findings regarding embedding quality, performance dynamics under concept drift, and the specific mechanisms driving classification behavior.

A comprehensive analysis of the temporal evolution of the model reveals an inverse drift phenomenon. As detailed in Section~\ref{sec:experiments}, the system exhibits a counter-intuitive trajectory where binary detection performance improves over time, while fine-grained attribution for the {Reconnaissance} class collapse.
Topological evidence clarifies this paradox. Initially, attack vectors such as {Mirai} and {Reconnaissance} form distinct, separable clusters in the latent space (Partition A). As time progresses, the network traffic shifts from low-volume scanning activities to high-intensity volumetric attacks. This transition makes the malicious traffic increasingly distinct from the {Benign} baseline, thereby simplifying the binary discrimination task. However, this same volumetric intensity creates a structural convergence between different attack classes, eroding the boundaries necessary for precise multiclass categorization.

The degradation in multiclass performance is not an indicator of model incapacity, but rather a reflection of the mimetic behaviour exhibited by evolving botnets. The topological shift observed in the experiments demonstrates that mature {Mirai} infections structurally mimic generic \ac{DoS} flooding attacks. In the final temporal partition, the {Mirai} embeddings migrate completely into the \ac{DoS} manifold, rendering them topologically indistinguishable. Beyond supervised intrusion detection, the efficacy of the unsupervised \ac{GNN-AE} model underscores the potential for autonomous network monitoring.

This paper introduced a \ac{GNN} pipeline that produces directly visualizable low-dimensional representations of \ac{IoT} network traffic through a dynamic multigraph architecture. These joint embeddings encode complex topological relationships and flow semantics which facilitate the immediate interpretation and monitoring of the network behavior. By projecting high-dimensional data into observable latent manifolds we establish an interpretable framework that also supports interoperability. Future research will utilize the direct movement of these clusters to preemptively identify adversarial shifts and ensure long-term robustness in dynamic environments. This transition toward visual topological monitoring provides a more sufficient approach to handle the fluid nature of behaviors of \ac{IoT} infrastructures.
\section{Acknowledgements}
This work was supported by the CyberFold project, funded by the European Union through the NextGenerationEU instrument (Recovery, Transformation, and Resilience Plan) [ETD202300129], managed by INCIBE; by the Research Grants HERMES [PID2023-152331OA-I00] and LATENTIA [PID2022-140786NB-C32], partially funded by the Ministry of Science and Innovation (MICIU) / State Research Agency (AEI) [10.13039/501100011033] and ERDF, FEDER / EU; partially by the Autonomous Community of Madrid (ELLIS Madrid Node) and the ELLIS Milan Node; and by the National Plan for NRRP Complementary Investments (PNC) in the call for the funding of research initiatives for technologies and innovative trajectories in the health [PNC0000003], under the project AdvaNced Technologies for Human-centrEd Medicine (ANTHEM).

\bibliographystyle{unsrtnat}
\bibliography{references}

\end{document}